\documentclass[submission,copyright,creativecommons]{eptcs}
\usepackage[english]{babel}
\providecommand{\event}{GaM 2016} 
\usepackage{breakurl}             
\usepackage{mathtools}
\usepackage{color}
\usepackage{tikz-cd}
\usepackage{mwe}
\usepackage[export]{adjustbox}
\usepackage{booktabs}

\title{A Graph Grammar for Modelling RNA Folding}
\author{Adane Letta Mamuye
\institute{School of Science and Technology\\
University of Camerino\\ 
Via del Bastione 1 62032\\
Camerino, Italy}
\email{adaneletta.mamuye@unicam.it}
\and
Emanuela Merelli
\institute{School of Science and Technology\\
University of Camerino\\
Via del Bastione 1 62032\\
Camerino, Italy}
\email{emanuela.merelli@unicam.it}
\and 
Luca Tesei
\institute{School of Science and Technology\\
University of Camerino\\
Via del Bastione 1 62032\\
Camerino, Italy}
\email{luca.tesei@unicam.it}
}
\def\titlerunning{A Graph Grammar for Modelling RNA Folding}
\def\authorrunning{A. L. Mamuye, E. Merelli \& L. Tesei}

\begin{document}
\maketitle

\begin{abstract}
We propose a new approach for modelling the process of RNA folding as a graph transformation guided by the global value of free energy. Since the folding process evolves towards a configuration in which the free energy is minimal, the global behaviour resembles the one of a self-adaptive system. Each RNA configuration is a graph and the evolution of configurations is constrained by precise rules that can be described by a graph grammar.
\end{abstract}

\section{Introduction}

Ribonucleic acid (RNA) is a molecule whose linear primary structure is composed of four different nucleotides: adenine \textbf{A}, guanine \textbf{G}, cytosine \textbf{C} and uracil \textbf{U}. RNA secondary structure is formed by the folding of the sequence of nucleotides through the crucial mechanism of base-pairing that allows three bonds: \textbf{A-U}, \textbf{G-C} and \textbf{G-U}. The three dimensional structure of the molecule is called RNA tertiary structure. RNA performs a variety of biological functions inside the cell such as protein synthesis, enzymatic catalysis and gene expression~\cite{mor-r}.  

RNA secondary structure can be determined by experimental techniques such as X-ray, crystallography and nuclear magnetic resonance, which are time consuming, expensive and in some cases infeasible. Consequently, for more than three decades numerous computational methods, among which comparative sequence analysis and dynamic programming algorithms, have been studied to predict the secondary structure starting from a given sequence of nucleotides. Comparative sequence analysis is the most reliable approach, while thermodynamics-based dynamic programming algorithms may be less accurate, but both are still computationally expensive. Thus, im\-pro\-ving the accuracy of predicting RNA secondary structure remains an ongoing challenge in computational biology. Moreover, it is still an open question at what extent the final structure assumed by RNA is determined by the minimal free energy with respect to the kinetic folding~\cite{fl_bey}. 

Since an RNA molecule exhibits an auto-regulative mechanism similar to the adaptability process of complex systems~\cite{mere_adp}, a new way of modelling the prediction of an RNA secondary structure can be investigated. The idea is to model the RNA folding by a graph-based approach that allows to represent the evolution of the structure step by step. Graph transformation can be viewed as an algebraic approach that creates a new graph from a given one by applying rewriting rules~\cite{cor-al}. In computational biology graph transformation has been applied in different contexts: Beck et al. showed how graph-rewriting algorithms can be employed to model one aspect of whole-organism morphogenesis~\cite{ben-gr}; the complexity of RNA tertiary structure motifs was encoded by a graph-grammar~\cite{st-gr}; and gene expression was simulated using a general purpose graph rewriting system~\cite{gr-gn}. Generally speaking, graph rewriting approaches is quite natural to use when modelling systems whose states have a network structure~\cite{bal-con}. 

In this work, an RNA secondary structure is represented as a graph and its folding evolution as a graph transformation in the folding space. The evolution continues until a configuration with minimum free energy is reached. Such transformations are expressed by derivations of a given {\em graph grammar}. In our case, each reached configuration delivers both local and global information: a representation of the current secondary structure and the corresponding free energy, respectively. We embed the dynamics given by graph transformation into the  $S[B]$ paradigm, a framework for modelling self-adaptive systems where $B$ is the local or behavioural component and $S$ is the global or structural one, both entangled in a unique model~\cite{em-tp}. 

The paper is organised as follows. In Section~\ref{sec:rna} we discuss the basics of RNA secondary structure. Section~\ref{sec:rgt} illustrates the RNA graph grammar and graph transformation. In Section~\ref{sec:rfe}, the RNA folding evolution is modelled as a self-adaptive system. Section~\ref{sec:implementation} introduces implementation issues while conclusions and future work are given in Section~\ref{sec:cl}. 

\section{RNA Secondary Structure} 
\label{sec:rna}

The single RNA strand, i.e.\ a sequence of nucleotides, is formed by bonds (interactions) between neighbour nucleotides, called the primary structure or the backbone. The primary structure folds to itself and leads to the formation of the secondary structure. The secondary structure is defined by weaker bonds (base-pair interactions) between two \emph{non-neighbour} nucleotides due to the Watson-Crick base-pairs (\mbox{\textbf{A-U}} and \textbf{G-C}) and the wobble base-pair \textbf{G-U}~\cite{hof-stad}. 

\begin{figure}[ht]
\centering
\includegraphics[scale=0.6]{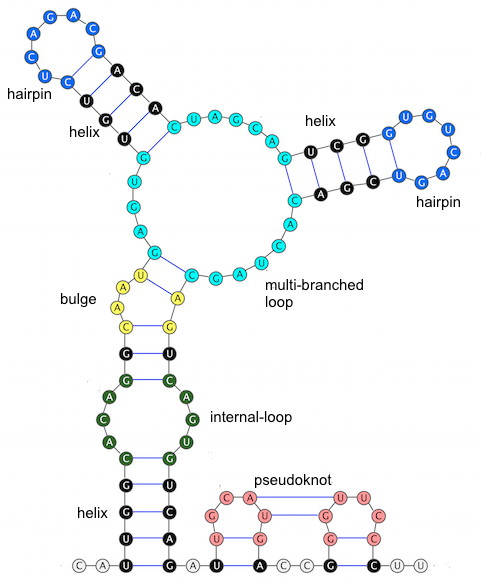}
\caption{An RNA secondary structure with examples of the five kinds of loops: hairpin (blue), bulge (yellow), helix (black), internal-loop (green) and multi-branched loop (cyan). Backbone bonds are depicted as grey lines while base-pairs as blue lines. An interaction among loops, i.e.\ a pseudoknot, is present in the bottom-right part (light pink) where two hairpin loops interact.} 
\label{fig:rnass}
 \end{figure}

The base-pair interactions produce the formation of different kinds of RNA structural elements called \emph{loops}, namely:
 \begin{itemize}
  \item {\em hairpin}: sequence of nucleotides enclosed by a single base-pair;
  \item {\em bulge}: two subsequent base-pairs with a sequence of nucleotides on one side; 
  \item {\em helix}: two or more subsequent base-pairs without a sequence of nucleotides in between (double helical region);
  \item {\em internal loop}: two subsequent base-pairs with a sequence of nucleotides on both sides; and
  \item {\em multi-branched loop}: three or more base-pairs that may be separated by sequences of nucleotides.
  \end{itemize}
Figure~\ref{fig:rnass} shows an RNA secondary structure in which the loops have been highlighted and labelled. Any RNA secondary structure can be of two types: pseudoknot free or pseudoknotted. A pseudoknot free structure is composed of a set of loops that do not interact each other, while a pseudoknot occurs when there is a base-pair interaction among loops. For instance, in Figure~\ref{fig:rnass}, the bottom-right part shows a pseudoknot in which two hairpin loops interact.  
  
\section{RNA Graph Grammar and Transformation} 
\label{sec:rgt}

An RNA secondary structure can be represented as a graph $G= (V, E)$ where $V$ is the set of vertices, labeled by the four nucleotides \textbf{A}, \textbf{C}, \textbf{G}, \textbf{U}, and $E$ is the set of bonds, both backbone and base-pair interactions. For using a graph rewriting approach we have to choose a particular graph-rewriting mechanism. Since RNA secondary structure can be viewed as a combination of basic structural elements, the algebraic approach proposed by Corradini et al.\ offers a significant advantage over the others~\cite{cor-al}. The main algebraic approaches are called DPO (double pushout) and SPO (single pushout). In DPO a direct derivation is given by two gluing diagrams, while in case of SPO a direct derivation is given by a single gluing diagram.   

\begin{figure}[ht!]
\begin{center} 
\includegraphics[scale=0.5]{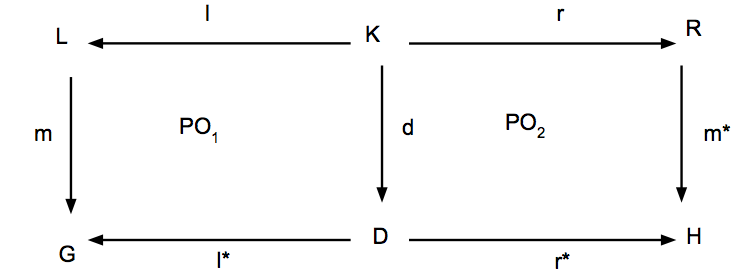} 

\ \\

\ \\

\ \\

\includegraphics[scale=0.48]{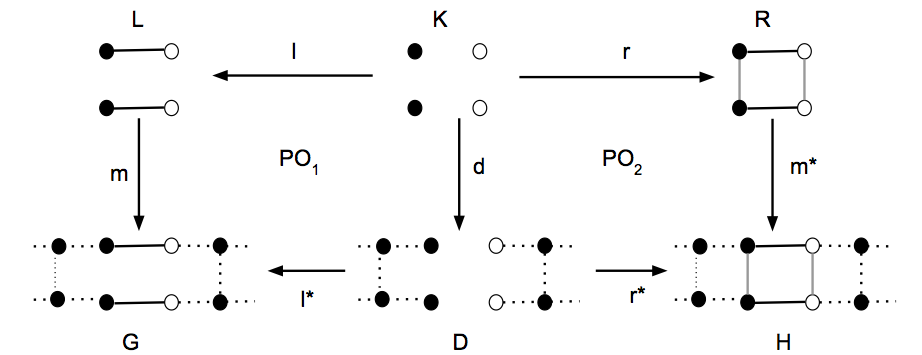}

\ \\

\end{center}
\caption{Above, the scheme of a production in the DPO approach. Below, a rewriting step corresponding to the application of a production (Helix Rule-1 of Table~\ref{pro}). Vertices of the same color (black or white) must form a base-pair, black edges represent backbone bonds and grey edges are base-pair bonds. Dotted lines indicate the possible presence of any kind of bond. The rewrite step proceeds as follows: $L$ is matched in $G$ and the intermediate graph $D$ is constructed by deleting edges of $L$ that are not in $K$. Then, $R$ is glued on $D$ to derive graph $H$.} 
\label{fig:dpo}
\end{figure}

The basic idea of DPO graph rewriting approaches is to consider a production $p: (L \leftarrow K \rightarrow R)$, where $L$ is a left-hand side (LHS) graph, the \emph{pattern} graph, $R$ is a right-hand side (RHS) graph, the \emph{rewrite} graph, and $K$ is an \emph{interface} graph that is embedded in $R$ and $L$. The interface graph is necessary to perform the rewriting step, but it is not affected by the step itself. Each graph production defines a total graph morphism and it must satisfy application conditions, called \emph{gluing conditions}. A production $p$ also determines which vertices and edges have to be preserved, deleted and created by its application. If a match $m$ identifies an occurrence of $L$ in a given graph $G$, then $G \xRightarrow{p,m} H$\footnote{For the sake of readability, the $m$ label can be dropped.} denotes a direct derivation step where $p$ is applied to $G$ to derive graph $H$. $H$ is obtained by replacing the occurrence of $L$ in $G$ by $R$ (see Figure~\ref{fig:dpo}). We use DPO to construct RNA secondary structures with gluing conditions, which in our case correspond to respecting the base-pairs constraints. 

We define an {\em RNA graph grammar} $G_{RNA} = (\{p: (L \leftarrow K \rightarrow R ) \}_{p \in P}, G_0)$ where $G_0$ is the start graph, i.e., in our case, the graph corresponding to the primary structure, and $P$ is the set of rewriting rules, as presented in Table~\ref{pro}. The application of the rewriting rules must take into account the biological constraints: 
\begin{itemize}
  \item base-pairing is possible only with \textbf{G-C}, \textbf{A-U} and \textbf{G-U} pairs;
  \item besides backbone bonds, each nucleotide can form a base-pair by interacting with at most one other nucleotide.
  \end{itemize}
These criteria are sufficient for ensuring that the rewriting rules are biologically admissible. As a result, if we are given a primary structure as a start graph $G_0$ and a set of productions $P$, we can derive a set of derivations such that, at each derivation step, one $p \in P$ is applied to the current graph. The folding space generated by the graph grammar is the set of all secondary structures that can be derived from the grammar starting from a primary structure. 
  
\begin{table}[ht]
\begin{center}
      \begin{tabular}{ccc}
         \toprule
                & 
         Rule-1 & 
         Rule-2 \\ 
         \midrule
         
         Hairpin &
         \includegraphics[valign=m,scale=0.4]{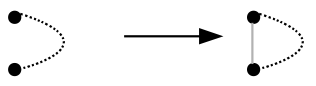} &
                        \                                                      \\
         \midrule
         
         Bulge-r & 
         \includegraphics[valign=m,scale=0.4]{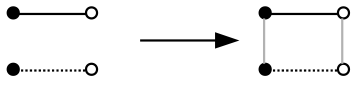} &
         \includegraphics[valign=m,scale=0.4]{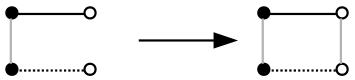} \\ 
         \midrule 
         
         Bulge-l &
         \includegraphics[valign=m,scale=0.4]{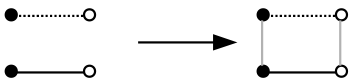} & 
         \includegraphics[valign=m,scale=0.4]{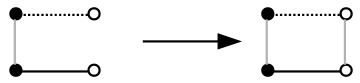} \\
         \midrule
         
         Helix & 
         \includegraphics[valign=m,scale=0.4]{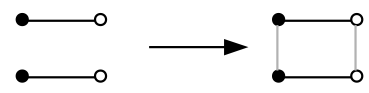} &
         \includegraphics[valign=m,scale=0.4]{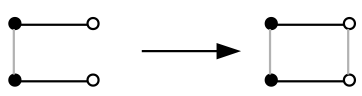} \\ 
         \midrule
         
         Internal-loop &
         \includegraphics[valign=m,scale=0.4]{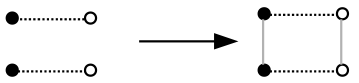} & 
         \includegraphics[valign=m,scale=0.4]{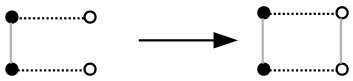} \\
         \midrule
         
         Multi-branched-loop &
         \includegraphics[valign=m,scale=0.4]{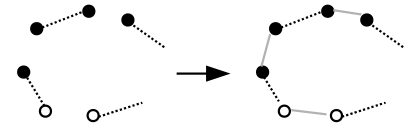} & 
         \includegraphics[valign=m,scale=0.4]{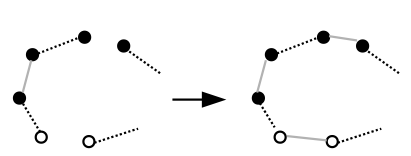} \\
         \bottomrule
      \end{tabular}
\end{center}
\caption{Rewriting rules for building structural elements (loops) of pseudoknot free RNA secondary structures. Only the LHS and the RHS of the productions are shown: the interface graph $K$ is the obvious one in each case (see for instance Figure~\ref{fig:dpo}). Black lines represent backbone bonds and vertices with the same color are meant to represent nucleotides that are admissible to form a base-pair. Dotted lines must match with a sequence of at least two (at least one in the Multi-branched-loop) backbone bonds.} 
\label{pro}
\end{table}

All the rewriting rules, given in Table~\ref{pro}, take an occurrence of compatible unpaired nucleotides and add an RNA loop which, according to the fact that loops are the basic structural elements of RNA secondary structures, is the basic primitive on which the graph grammar is defined. We present only the LHS graph and the RHS graph of the productions because the interface graph is the obvious one in each case, as shown in~Figure~\ref{fig:dpo}. Let us briefly explain the rules by starting from the easiest ones. The Helix rules, Rule-1 and Rule-2, model the formation of an helix sub-structure. Helix Rule-1 assumes four vertices (nucleotides) in the LHS graph with existing backbone pairs. Black lines represent backbone bonds and vertices with the same color are meant to represent nucleotides that are admissible to form a base-pair. Helix Rule-2 assumes the existence of the same nucleotides plus one base-pair. This production treats the cases in which at least one helix loop is already present and another one is added. If the match criteria are fulfilled, then the RNA helix sub-structure will be glued to the graph, i.e.\ two (or one) base-pair bonds, represented by grey lines, are added. Following the same scheme (also regarding the difference between Rule-1 and Rule-2, apart from the single Hairpin rule), the rest of the rewriting rules glues hairpins, bulges, internal loops and multi-branched loops on a given secondary structure $G$. Note that the dotted lines in the rules for hairpin, bulge, internal-loop refer to the existence of one or more unpaired nucleotides in the backbone relation, while the dotted lines in the rules for the multi-branched-loop refer to the existence of zero or more unpaired nucleotides in the backbone relation. 

\begin{figure}[ht]
   \centering
\includegraphics[width=\textwidth]{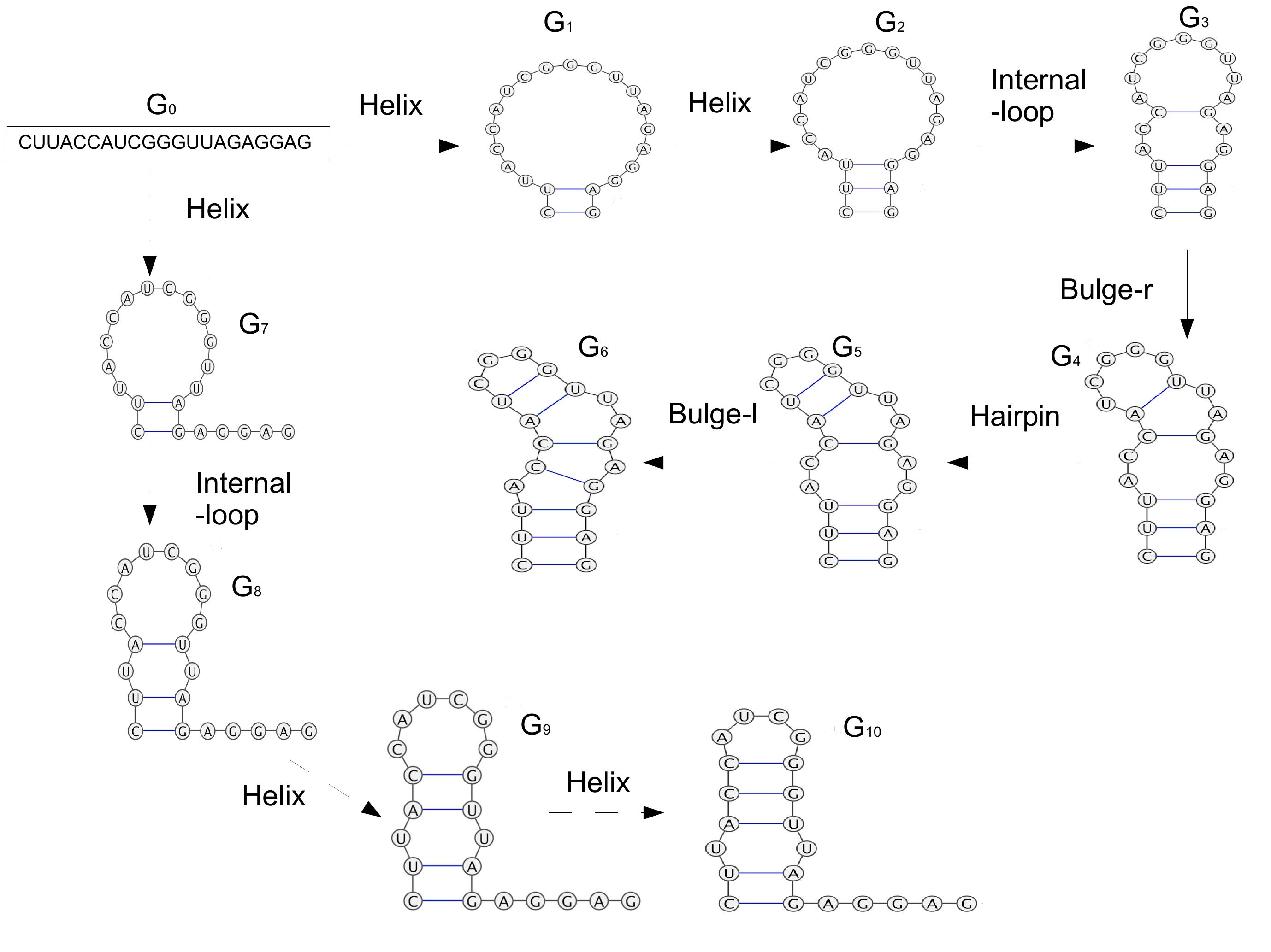} 
\caption{Two different derivations of RNA secondary structures starting from a primary structure $G_0$, which can also be seen as a graph where only the edges representing backbone bonds among the nucleotides are present. The first derivation, depicted with solid line arrows, is $G_0 \xRightarrow{\mathrm{Helix \; Rule-1}} G_1 \xRightarrow{\mathrm{Helix\; Rule-2}} G_2 \xRightarrow{\mathrm{Internal-loop\; Rule-2}} G_3 \xRightarrow{\mathrm{Bulge-r \; Rule-2}} G_4 \xRightarrow{\mathrm{Hairpin}} G_5 \xRightarrow{\mathrm{Bulge-l \; Rule-2}} G_6$. The second derivation, depicted with dashed line arrows, is $G_0 \xRightarrow{\mathrm{Helix \; Rule-1}} G_7 \xRightarrow{\mathrm{Internal-loop \; Rule-2}} G_8 \xRightarrow{\mathrm{Helix \; Rule-2}} G_9 \xRightarrow{\mathrm{Helix \; Rule-2}}G_{10}$.}  
\label{fig:der}
 \end{figure}

Based on $G_{RNA}$, a possibly empty derivation starting from an RNA primary structure $G_0$ and passing through intermediate RNA secondary structures $G_1, \ldots, G_n$, $n \geq 0$, is written $G_0 \xRightarrow{p_1} G_1 \xRightarrow{p_2} \cdots \xRightarrow{p_{n}} G_n$, where $p_i \in P$, $i>0$, and is possibly abbreviated with $G_0 \xRightarrow{*} G_n$. As a result, the language generated by $G_{RNA}$ is the set of all RNA secondary structures $G_n$, $n \geq 0$, that are derived with the grammar starting from any primary structure $G_0$. Two examples of derivations are given in~Figure~\ref{fig:der}. 

The two derivations of~Figure~\ref{fig:der} model two possible folding transformations of the same RNA molecule of $21$ nucleotides. The most commonly used RNA secondary structure prediction method is based on free energy minimization. Let us now add the free energy information to the secondary structures depicted in Figure~\ref{fig:der}. The \texttt{RNAeval} web server~\cite{vina-rna}, which gives a detailed thermodynamic description according to the loop-based energy model, can be used to compute the free energy $e(G)$ of each structure $G$. In the first derivation (with solid line arrows) $e(G_1) = 4.80$, $e(G_2)= 2.90$, $e(G_3) = 3.60$, $e(G_4) = 7.60$, $e(G_5)=6.70$ and $e(G_6)= 7.20$ (all values are in {\em kcal/mol}). Note that graph rewriting cannot transform $G_6$ further. The free energies in the second derivation (with dashed line arrows) are $e(G_7)= 2.60$, $e(G_8)= 3.90$, $e(G_9) = 0.70$, $e(G_{10}) = -2.80$. The lowest free energy structure of the two transformations are $e(G_3)= 3.60$ and $e(G_{10})=-2.80$. Considering the plot of the free energies at each step of the derivations, we can say that $e(G_3)= 3.60$ is a local minimum of the plot of the first derivation and $e(G_{10})=-2.80$ is a local minimum of the plot of the second one. Following this procedure, in a first naive brute-force approach, we can derive all the possible derivations and we can select, at the end, the optimal structure. For instance, for the given sequence, the optimal secondary structure predicted by the \texttt{RNAfold} web server~\cite{vina-rna} has a structure similar to our $G_{10}$ and a minimum free energy equal to $-2.80$ kcal/mol, which is equal to the free energy of $G_{10}$.

\section{RNA Folding Process as a Self-adaptive System} 
\label{sec:rfe}

\begin{figure}[ht]
\centering
\includegraphics[scale=0.38]{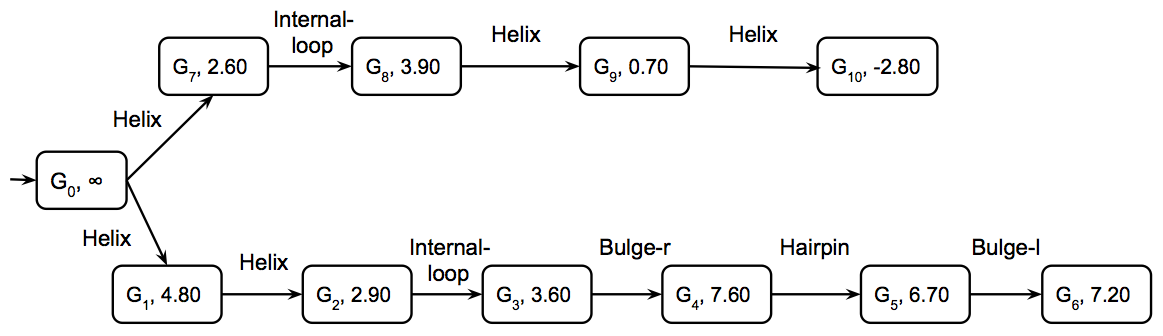} 
\caption{Part of the LTS representing all possible applications of the productions of the given graph grammar to the primary structure of 21 nucleotides $G_0$ of Figure~\ref{fig:der}. Besides the graph, each state of the LTS contains also the free energy value associated to the secondary structure represented by the graph. If the graph cannot be further transformed, the state has no outgoing transitions. Only the two derivations of Figure~\ref{fig:der} are depicted, neglecting other derivations and possible branchings along these two.}  
\label{fig:lts}
 \end{figure}

RNA primary structure folds until it reaches a stable folding configuration. In this folding process, due to non-determinism, many possible secondary structure configurations can be generated. We introduced a graph grammar for generating all possible RNA secondary structures by taking into account a set of production rules. In other words, for each primary structure $G_0$ a Labelled Transition System (LTS) can be defined in which the initial state is the graph $G_0$ and the other states are all the possible graphs derivable from it using the graph rewriting rules. Part of such an LTS, containing only the two derivations presented in~Figure~\ref{fig:der}, is shown in Figure~\ref{fig:lts}. 

Each state of the LTS contains a graph corresponding to a secondary structure and its corresponding free energy value given in kcal/mol. The transitions exactly correspond to the application of one production $p$ to their source state graph obtaining their target state graph. 

At each state in the LTS the value of the free energy is called the \emph{observable} value. In our interpretation of the RNA folding process as a self-adaptive system, these observables represent the \emph{environment} in which the system is immersed. Changes in these values, i.e.\ changes in the environment, are perceived by the system and are used to possibly trigger adaptation.  

In general, self-adaptive systems are able to modify their own behaviour according to their current configuration and the perception of the environment in which they operate. This feature is formalized by the $S[B]$ paradigm \cite{em-tp,mere_adp,mr-tp}. An $S[B]$ model, following the paradigm, comprises two coupled levels: the behavioural level $B$, which describes the admissible dynamics of the system and the structural level $S$, accounting for the invariant features of the system and regulating its entangled behaviour as a whole. 

The behavioural level can be defined as a finite state machine of the form $B=(Q, q_0, \rightarrow_B)$ where $Q$ is a set of states, $q_0$ is the initial state and $\rightarrow_B$ is the transition relation. The structural level can also be modelled as a finite state machine $S = (W,w_0,\mathcal{O},\rightarrow_S,\mathcal{C})$ where $W$ is a set of states, $w_0$ is the initial state, $\mathcal{O}$ is an observation function that gives a value for each state of $B$, $\rightarrow_S$ is the transition relation and $\mathcal{C}$ is a function that associate a formula, representing a set of constraints, to every state in $W$ and to every transition $(w_i,w_j) \in \rightarrow_S$. The constraints $\mathcal{C}(w)$ are over the observables and the current configuration of the whole $S[B]$ model and are meant to represent invariant conditions that must be fulfilled while the system is in the steady situation represented by the $S$ state $w$. The constraints $\mathcal{C}((w_i,w_j))$ are over the same information, but they must be fulfilled during an adaptation phase that starts in the steady state $w_i$ and ends in the steady state $w_j$. Such constraints are meant to be safety conditions that may be needed to ensure during adaptation. However, if a completely unconstrained adaptation is needed they can be set to $\mathit{true}$.

In simple terms the adaptation model of $S[B]$ can be viewed as a closed-loop system where $B$ is the plant and $S$ is the controller. Let us informally describe the semantics. At each time instant the $S[B]$ model is in a state $w[q]$ such that $w \in W$ and $q\in Q$. If the current constraint $\mathcal{C}(w)$ is satisfied by the current observables $\mathcal{O}(q)$ and by the current configuration of the two state machines $S$ and $B$, then the whole $S[B]$ model is in a steady (non-adapting) state. The $S[B]$ model can evolve from $w[q]$ to any $w[q']$, for some $q' \in Q$ such that $q \rightarrow_B q'$, if the current constraint $\mathcal{C}(w)$ continues to hold in $w[q']$. The actual choice of $q'$ may be influenced by the instantiation of some variables in $\mathcal{C}(w)$. If, instead, the invariant condition $\mathcal{C}(w)$ cannot be fulfilled by any successor state $q' \in Q$, then the $S[B]$ model starts an adaptation phase in which it searches a new $S$ state $w'$, successor of $w$ in $S$, such that its invariant condition $\mathcal{C}(w')$ can be satisfied. In the adaptation phase $B$ is no more constrained by $S$, apart from the safety constraints $\mathcal{C}((w,w'))$ that must hold during the whole adaptation phase. Adaptation terminates successfully when $B$ ends up in a state that fulfills the new global situation represented by one of the admissible $S$ states $w'$ that are successors of $w$.

\begin{figure}[ht]
\centering
\includegraphics[scale=0.4]{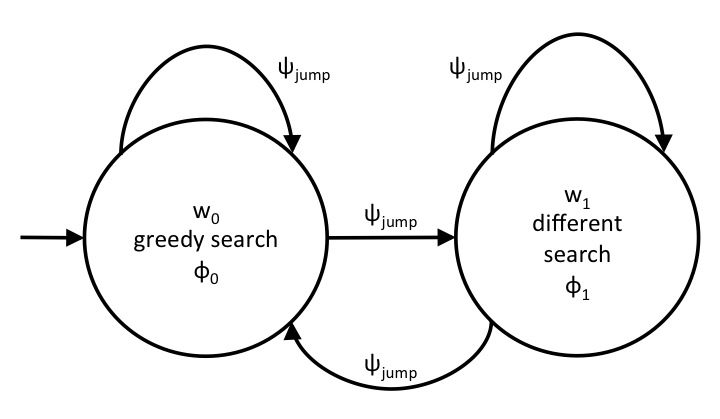} 
\caption{A possible $S$ level of the $S[B]$ model for the RNA folding process.}  \label{fig:s}
 \end{figure}

Let us use the LTS defined in Figure~\ref{fig:lts} as the $B$ level of an $S[B]$ model for the RNA folding process. Formally, to obtain the $B$ level from the LTS we have to forget all the transition labels. For the $S$ level we use the state machine depicted in Figure~\ref{fig:s}. Each state $w_i, i = \{0,1\}$, has an associated constraint $\phi_i = \mathcal{C}(w_i)$. Each transition in $\rightarrow_S$ has an associated constraint $\psi_\mathrm{jump}$ that, for the sake of simplicity, we initially suppose set to $\mathit{true}$. 
The initial state is $w_0$ and its associated constraint is as follows:
$$
\mathcal{C}(w_0) = \phi_0 = \exists q_\mathrm{next} \in \mathrm{next}(q) \colon \mathcal{O}(q_\mathrm{next}) \leq \mathcal{O}(q) \wedge (\forall q' \in \mathrm{next}(q) \colon \mathcal{O}(q_\mathrm{next}) \leq \mathcal{O}(q'))
$$

where $q$ is the current state of the $B$ level, i.e.\ the current RNA secondary structure, $\mathcal{O}(q)$ is the observable value associated to $q$, in our case the free energy associated to the RNA secondary structure at $q$, and $\mathrm{next}(q)$ is a function giving the successors states of $q$ in $B$, i.e.\ all the possible graphs that can be reached by applying one graph transformation $p$ of the defined graph grammar.  

The constraint $\phi_0$, read as an invariant, expresses the fact that there is a possible evolution of the current RNA secondary structure at state $q$ into an RNA secondary structure at state $q_\mathrm{next}$ such that the free energy is less than the current one and it is the lowest among all the possible alternatives. This clearly defines a \emph{greedy} dynamics for the evolution of the $S[B]$ model, towards the closest (in terms of number of applications of productions $p$ of the graph grammar) RNA secondary structure whose free energy is a local minimum (in the space of reachable states of the $B$ level). 

Let us consider, for instance, our example LTS in Figure~\ref{fig:lts}. At the beginning the $S[B]$ system is in state $w_0[G_0]$\footnote{With an abuse of notation we use $G_i$ to denote the state of the LTS whose associated graph is $G_i$.}. The observable, i.e.\ the free energy, of the current state is $\mathcal{O}(G_0) = +\infty$. The constraint of the initial $S$ state, $\mathcal{C}(w_0) = \phi_0$, can be satisfied by a proper instantiation of $q_\mathrm{next}$ choosing between the two successor states, one associated with graph $G_1$ and one with graph $G_7$. The only possible choice to satisfy $\phi_0$ is to instantiate $q_\mathrm{next}$ with the state associated to $G_7$. Thus, the $S[B]$ model moves to state $w_0[G_7]$. At this point, in state $w_0[G_7]$, the global constraint $\phi_0$ cannot be satisfied anymore because the only available successor state of $G_7$ has a free energy that is greater than the current one. As we mentioned before, this state is a local minimum along the sequence of states form $G_0$ to $G_{10}$. Following the given semantics for $S[B]$, in state $w_0[G_7]$ an adaptation phase starts. In the state machine for the $S$ level in Figure~\ref{fig:s} there are two possible successor steady states of $w_0$, i.e.\ $w_0$ itself and $w_1$. The constraint formula $\psi_\mathrm{jump}$ associated to both the transitions in $S$ is $\mathit{true}$, thus, during adaptation the $S[B]$ model can explore the whole state space of $B$ without restrictions. In our example, the only possible exploration is to continue towards state $G_8$. Note that in $G_8$ the constraint $\phi_0$ is again satisfied, thus the $S[B]$ model can stop adaptation at the steady state $w_0[G_8]$. Continuing the execution, the $S[B]$ model will continue in the steady state $w_0$ until it reaches $G_{10}$, where it will stop. Note that, during the execution, the $S[B]$ model outputs its traces, thus the lower local minimum found during its functioning can be recovered, in this case the observable value associated to the last output. 

The state $w_1$ in Figure~\ref{fig:s} is unspecified as it is left as a future work. We can instantiate ``different search'' with other searching strategies imported from the domain of non-linear optimization or we can invent other strategies based on biological information on the RNA folding. Moreover, we can add other $S$ states for trying different strategies at the same time. Finally, we should also enrich the $B$ state machine, i.e.\ the LTS, by adding the reverse transition of $\rightarrow_B$ in order to permit backtracking. The reverse transition should be easily derivable by inverting the production rules of the graph grammar. The use of the reverse transition could be allowed only during adaptation phases or at any time of the execution of the $S[B]$ model by associating suitable constraints to the $S$ states $w_i$. 

\section{Implementation Issues}
\label{sec:implementation}

For testing our approach, we used the \texttt{GROOVE}~\cite{groove} simulator. Note that \texttt{GROOVE} does not support DPO, but in general it is possible to obtain the same behaviour by using SPO with suitable restricted application conditions~\cite{cor2}. Thus, we used restricted SPO for generating the LTS of the graph transformations of the primary structure with $21$ nucleotides given in~Figure~\ref{fig:der}. Even for this short sequence, the size of the state space was big. Thus, attempting to explore the complete folding space, as our approach suggests, is computationally expensive. 

In general, the folding space for the RNA secondary structure starting with a sequence of $n$ nucleotides has approximatively $1.8^n$ possible states, as defined by Equations (3-7) in~\cite{zuker_rna}. Different strategies were developed to handle scalability issues. Dynamic programming (DP), the most investigated approach, implicitly explores the RNA secondary structure space to find the lowest free energy structure without explicitly generating all possible structures. The Zuker and Stiegler thermodynamic model is considered as a benchmark for computational RNA structure prediction~\cite{zuk_opt}. According to this model, the free energies of RNA secondary structures can be recursively calculated, via DP, as the sum of the energy contribution of its loops. A practical strategy to reduce the complexity is to use a stochastic simulation that can be described as a continuous time Markov process~\cite{fl_bey}.

To mitigate the computational effort required by our approach, the number of states, i.e.\ the possible secondary structures, must be reduced. It is our aim, as future work, to define and exploit a partition function for the RNA folding space, then to compute the base-pairing probability of each possible base-pair of a given sequence and to improve the identification of the replacement graph with the most probable base-pairs.   

\section{Conclusions and Future Work} 
\label{sec:cl}
We have started to devise a new approach for modelling the RNA folding evolution as a self adaptive system within the $S[B]$ paradigm. In doing so, we have considered graph transformation as the main technique to naturally define the folding evolution of an RNA strand. The behavioural level $B$ of the defined $S[B]$ model is given by an LTS whose states contains graphs representing secondary structure and whose transitions are derived using a given graph grammar. The structural level $S$ is a finite automaton that accounts for monitoring the adaptability process that evolves towards an RNA secondary structure with a minimum free energy. 

As future work, we plan to expand our approach to treat also pseudoknotted secondary structures. Moreover, as we outlined in Section~\ref{sec:rfe}, we plan to complete the definition of the behavioural and structural levels in order to implement searching strategies typical of non-linear optimization and, possibly, other smart strategies based on the biological knowledge of the domain. Finally, to mitigate the computational effort required by exploring the whole folding space, besides the method outlined at the end of Section~\ref{sec:implementation}, we will consider the natural topological classification of RNA structures in terms of irreducible components that are embedable in surfaces of fixed genus~\cite{reidys}. 

\section*{Acknowledgments}
We acknowledge the financial support of the Future and Emerging
Technologies (FET) programme within the Seventh Framework Programme (FP7)
for Research of the European Commission, under the FET-Proactive grant
agreement TOPDRIM, number FP7-ICT-318121.

\bibliographystyle{eptcs}
\bibliography{generic}

\end{document}